\documentclass{aa}

\usepackage{graphicx}
\usepackage{txfonts}
\usepackage{natbib}
\usepackage{xcolor}
\usepackage{xspace}
\usepackage[colorlinks=true,linkcolor=blue,citecolor=blue]{hyperref}
\usepackage{etoolbox}

\makeatletter

\patchcmd{\NAT@citex}
  {\@citea\NAT@hyper@{%
     \NAT@nmfmt{\NAT@nm}%
     \hyper@natlinkbreak{\NAT@aysep\NAT@spacechar}{\@citeb\@extra@b@citeb}%
     \NAT@date}}
  {\@citea\NAT@nmfmt{\NAT@nm}%
   \NAT@aysep\NAT@spacechar\NAT@hyper@{\NAT@date}}{}{}

\patchcmd{\NAT@citex}
  {\@citea\NAT@hyper@{%
     \NAT@nmfmt{\NAT@nm}%
     \hyper@natlinkbreak{\NAT@spacechar\NAT@@open\if*#1*\else#1\NAT@spacechar\fi}%
       {\@citeb\@extra@b@citeb}%
     \NAT@date}}
  {\@citea\NAT@nmfmt{\NAT@nm}%
   \NAT@spacechar\NAT@@open\if*#1*\else#1\NAT@spacechar\fi\NAT@hyper@{\NAT@date}}
  {}{}

\makeatother

\newcommand{\hlink}[1]{\url{http://#1}\xspace}

\newcommand{\rfig}[1]{Fig.~\ref{#1}}
\newcommand{\rfigs}[1]{Figs.~\ref{#1}}
\newcommand{\req}[1]{Eq.~\ref{#1}}

\newcommand{\rtab}[1]{Table \ref{#1}}

\newcommand{\rapp}[1]{Appendix \ref{#1}}

\newcommand{\rsec}[1]{section \ref{#1}}

\newcommand{\herschel}{{\it Herschel}\xspace}

\newcommand{\jwst}{{\it JWST}\xspace}

\newcommand{\um}{\mu{\rm m}}

\newcommand{\Jykms}{{\rm Jy.km/s}}

\newcommand{\sfr}{{\rm SFR}}

\newcommand{\ssfr}{{\rm sSFR}}
\newcommand{\lir}{L_{\rm IR}}
\newcommand{\lfir}{L_{\rm FIR}}

\newcommand{\cii}{{\rm [\ion{C}{II}]}\xspace}
\newcommand{\nii}{{\rm [\ion{N}{II}]}\xspace}
\newcommand{\ciion}{{\rm [\ion{C}{II}]_{ion}}}
\newcommand{\fion}{f_{\rm \cii,ion}}
\newcommand{\fism}{f_{\rm IR,ISM}}
\newcommand{\niion}{{\rm [\ion{N}{II}]_{ion}}}
\newcommand{\lcii}{L_{[\ion{C}{ii}]}}

\newcommand{\lsun}{L_\odot}
\newcommand{\msun}{{\rm M}_\odot}

\newcommand{\kpc}{{\rm kpc}}

\newcommand{\Gyr}{{\rm Gyr}}
\newcommand{\Myr}{{\rm Myr}}
\newcommand{\yr}{{\rm yr}}
\newcommand{\dex}{{\rm dex}}
\newcommand{\mstar}{M_\ast}

\newcommand{\tdust}{T_{\rm dust}}

\newcommand{\mean}[1]{\left<#1\right>}

\newcommand{\kelvin}{{\rm K}}
\newcommand{\hyde}{{\it Hyde}\xspace}
\newcommand{\jekyll}{{\it Jekyll}\xspace}

\newcommand{\halpha}{${\rm H}_\alpha$\xspace}

\defcitealias{schreiber2015}{S15}
\defcitealias{schreiber2018}{S18}
\defcitealias{chary2001}{CE01}
\defcitealias{dale2002}{DH02}
\defcitealias{galliano2011}{G11}
\defcitealias{draine2007}{DL07}

\newcommand*\dd{\ensuremath{d}}

\begin{document}

\title{A low [CII]/[NII] ratio in the center of a massive galaxy at $z=3.7$: witnessing the transition to quiescence at high-redshift?}
\titlerunning{A low [CII]/[NII] ratio in the center of a massive galaxy at $z=3.7$}

\author{C.~Schreiber\inst{1}
  \and K.~Glazebrook\inst{2}
  \and C.~Papovich\inst{3}
  \and T.~D\'iaz-Santos\inst{4}
  \and A.~Verma\inst{1}
  \and D.~Elbaz\inst{5}
  \and G.~G.~Kacprzak\inst{2}
  \and T.~Nanayakkara\inst{6}
  \and P.~Oesch\inst{7,8}
  \and M.~Pannella\inst{9}
  \and L.~Spitler\inst{10,11}
  \and C.~Straatman\inst{12}
  \and K.-V.~Tran\inst{13,14}
  \and T.~Wang\inst{15,16}
}

\institute{
    Astrophysics, Department of Physics, Keble Road, Oxford OX1 3RH, United Kingdom \\
    \email{cschreib@orange.fr} 
    \and Centre for Astrophysics and Supercomputing, Swinburne University of Technology, Hawthorn, VIC 3122, Australia 
    \and George P.~and Cynthia W.~Mitchell Institute for Fundamental Physics and Astronomy, Department of Physics and Astronomy, Texas A\&M University, College Station, TX 77843, USA 
    \and N\'ucleo de Astronom\'ia, Facultad de Ingenier\'ia y Ciencias. Universidad Diego Portales, Ej\'ercito Libertador 441, Santiago, 8320000, Chile. 
    \and AIM-Paris-Saclay, CEA/DSM/Irfu -- CNRS -- Universit\'e Paris Diderot, CEA-Saclay, pt courrier 131, 91191 Gif-sur-Yvette, France 
    \and Leiden Observatory, Leiden University, NL-2300 RA Leiden, The Netherlands 
    \and Department of Astronomy, University of Geneva, 51 Ch.~des Maillettes, 1290 Versoix, Switzerland 
    \and International Associate, Cosmic Dawn Center (DAWN) 
    \and Faculty of Physics, Ludwig-Maximilians Universit\"at, Scheinerstr.~1, 81679 Munich, Germany 
    \and Research Centre for Astronomy, Astrophysics \& Astrophotonics, Macquarie University, Sydney, NSW 2109, Australia 
    \and Department of Physics \& Astronomy, Macquarie University, Sydney, NSW 2109, Australia 
    \and Max-Planck Institut f\"ur Astronomie, K\"onigstuhl 17, D-69117, Heidelberg, Germany 
    \and Australia Telescope National Facility, CSIRO Astronomy and Space Science, PO Box 76, Epping, NSW 1710, Australia 
    \and School of Physics, University of New South Wales, Sydney, NSW 2052, Australia 
    \and Institute of Astronomy, The University of Tokyo, Osawa, Mitaka, Tokyo 181-0015, Japan 
    \and National Astronomical Observatory of Japan, Mitaka, Tokyo 181-8588, Japan 
}

\date{Received 5 August 2019; accepted 26 November 2020}
{
\abstract {Understanding the process of quenching is one of the major open questions in galaxy evolution, and crucial insights may be obtained by studying quenched galaxies at high redshifts, at epochs when the Universe and the galaxies were younger and simpler to model. However, establishing the degree of quiescence in high redshift galaxies is a challenging task. One notable example is \hyde, a recently discovered galaxy at $z_{\rm spec}=3.709$. As compact ($r_{1/2}\sim0.5\,\kpc$) and massive ($\mstar\sim10^{11}\,\msun$) as its quenched neighbor \jekyll, it is also extremely obscured yet only moderately luminous in the sub-millimeter. Panchromatic modeling suggested it could be the first galaxy found in transition to quenching at $z>3$, however the data were also consistent with a broad range of star-formation activity, from fully quenched to moderate star formation rates (SFR) in the lower scatter of the galaxy main-sequence. Here, we describe Atacama Large Millimeter Array (ALMA) observations of the $\cii$ $157\,\um$ and $\nii$ $205\,\um$ far-infrared emission lines. The \cii emission within the half-light radius is dominated by ionized gas, while the outskirts are dominated by photo-dissociation regions or neutral gas. This suggests that the ionization in the center is not primarily powered by on-going star formation, and could come instead from remnant stellar populations formed in an older burst, or from a moderate active galactic nucleus (AGN). Accounting for this information in the multi-wavelength modeling provides a tighter constraint on the star formation rate of $\sfr=50^{+24}_{-18}\,\msun/\yr$. This rules out fully quenched solutions, and favors $\sfr$s more than factor of two lower than expected for a main-sequence galaxy, confirming the nature of \hyde as a transition galaxy. Theses results suggest that quenching happens from inside-out, and starts before the galaxy expels or consumes all its gas reservoirs. Similar observations of a sample of massive and obscured galaxies would determine whether this is an isolated case or the norm for quenching at high-redshift.
}


\maketitle

\section{Introduction}

Massive galaxies in the low-redshift Universe are observed to be mostly quiescent, with current star formation rates less then 1\% of their past average (e.g., \citealt{pasquali2006}). But how galaxies quench their star formation and turn into massive, red early-type galaxies is one of the key unresolved questions of galaxy evolution.

While a number of mechanisms have been proposed in the literature to cause galaxies to stop, reduce, or prevent star formation -- including black hole feedback, strong outflows, or gas stripping (e.g., \citealt{silk1998,birnboim2003,croton2006,gabor2012,martig2009,foersterschreiber2014,genzel2014,peng2015}), it is still unknown which of these mechanism is most important to explain emergence of quenched galaxies. Crucial insight can be gained by studying quiescent galaxies at higher redshift, where the available time for feedback processes to act is shorter, limiting the range of possible mechanisms. Recent work has shown that massive quiescent galaxies already existed as early as $z=4$ (e.g. \citealt{labbe2005,kriek2009,gobat2012,merlin2018,schreiber2018-b,belli2019}), at an epoch where state-of-the-art numerical simulations predict all galaxies to be forming stars \citep[e.g.,][]{wellons2015,dave2016}. These galaxies must have had massive star-forming progenitors at even higher redshifts, and because the age of the Universe was then comparable to their estimated stellar ages (of order a billion years or less), they must have quenched shortly before being observed. The actor of this abrupt quiescence may therefore be easier to identify than in lower redshift objects, which are seen after several billion years of passive evolution.

An interesting case is that of the most distant known quiescent galaxy to date, at $z_{\rm spec}=3.715$ \citep{glazebrook2017}. Detection of sub-millimeter emission toward this object \citep{simpson2017-a} was later found to arise from to a neighboring extremely obscured galaxy at the same redshift (\citealt{schreiber2018}, hereafter \citetalias{schreiber2018}). This was demonstrated by the detection with ALMA of the \cii line blueshifted by $550\,{\rm km/s}$ compared to the Balmer absorption lines of the quiescent galaxy, and most convincingly by the improved spatial resolution ($0.4\arcsec$) and depth of the sub-millimeter imaging, showing that both the \cii and dust emissions are produced by a separate, rotating galaxy, mostly unresolved ($0.1\arcsec$ radius) and located $0.5\arcsec$ away from the quiescent galaxy. The pair was dubbed ``\jekyll and \hyde'', with \jekyll being the quiescent galaxy, and \hyde the obscured galaxy.

In \citetalias{schreiber2018}, we performed an extensive analysis of the rich multi-wavelength data at hand to understand the physical conditions in the obscured galaxy \hyde. We showed that, despite its extreme obscuration and sub-mm detection, this galaxy appears to form stars at a relatively slow pace, with a star-formation rate $\sfr<100\,\msun/\yr$ and a stellar mass $\mstar\simeq10^{11}\,\msun$. Given the strong obscuration ($A_V\simeq3$) and the galaxy's large stellar mass, the infrared luminosity is in fact low enough ($\lir\simeq10^{12}\,\lsun$) that it may be entirely powered by intermediate-age stars, such that its current $\sfr$ (averaged over the last $10\,\Myr$) could be as low as zero. Although these data could not exclude that \hyde is simply a normal galaxy in the lower envelope of the main sequence, the possibility of it being quenched or in transition to quenching is particularly interesting. Indeed, given that this galaxy is still obscured and is therefore likely to contain substantial gas reservoirs, this would be at odds with a number of proposed quenching mechanisms which require full removal or consumption of the gas reservoirs prior to quenching.

This surprising conclusion is nevertheless independently supported by a number of pieces of evidence, first reported in \citetalias{schreiber2018} and summarized here for the reader's convenience: a) its dust temperature ($\tdust\simeq30\,\kelvin$) is almost $10\,\kelvin$ lower than the average for $z\sim4$ galaxies \citep{schreiber2018-a}, which suggests a softer-than-average radiation field, b) its compact dust continuum size of about $0.5\,\kpc$ is smaller than the rest-ultraviolet size of all $z\sim4$ star-forming galaxies in the same field, and is instead similar to that of quiescent galaxies \citep{straatman2015}, and c) its $\lcii/\lfir$ ratio is abnormally low for a galaxy of this luminosity, which we showed could be explained by a recent truncation of star formation, although other explanations for the observed $\cii$ deficit could not be excluded. Considered independently, none of these facts is conclusive or extremely unusual, and indeed there are other examples of galaxies which share at least one of these properties (e.g., similar compactness, dust temperature, or $\lcii/\lfir$ ratio). It is, however, the sum of these facts which suggests an abnormal process is at play in this galaxy; to our knowledge, such a combination of observables is unique among distant massive galaxies. Although all these signs may point toward \hyde being caught in a very specific phase, possibly in transition to quiescence, none of the data available at the time allowed us to prove this conclusively. In light of the results presented in this paper, we describe these points further in our discussion.

In this paper, we exploit new spectroscopic data from the Atacama Large Millimeter Array (ALMA) to address this question, combining observations of the \cii and \nii ($205.178\,\um$) emission lines. \nii, owing to its high ionization potential ($14.5\,{\rm eV}$), is exclusively found in ionized gas regions, where it can be found alongside \cii with an almost constant line ratio $\ciion/\niion\sim3$ (see \citealt{oberst2006}). In contrast, photo-dissociation regions (PDRs) surrounding stellar birthclouds are a neutral medium that emits \cii but no \nii. The observed $\cii/\nii$ ratio can therefore be used as a probe of the fraction of the \cii-emitting gas which is associated with ionized gas. While star-forming regions will typically be comprised of both neutral and ionized gas, the former is found to dominate the \cii emission in star-forming galaxies \citep{pavesi2016,diaz-santos2017}. In contrast, nearby early-type galaxies, which are forming stars at much lower rates, were shown to have a significantly larger fraction of their \cii associated with ionized gas \citep{lapham2017}. This implies that the ionized component of \cii is mostly not associated with star-formation, hence that the $\cii/\nii$ ratio can be used as an independent tracer of star formation activity, which we apply here to the galaxy \hyde.

The structure of this paper is as follows. In \rsec{SEC:data} we describe the new ALMA observations and detections, in \rsec{SEC:model} we describe the method used to analyze them, in \rsec{SEC:results} we describe our results and the ionization state of the gas, and use these observations to refine our estimate of the current $\sfr$ in \hyde. In \rsec{SEC:discussion} we discuss how other observables support these results, and we finally conclude in \rsec{SEC:conclusion}.

In the following, we assumed a $\Lambda$CDM cosmology with $H_0 = 70\ {\rm km}\,{\rm s}^{-1} {\rm Mpc}^{-1}$, $\Omega_{\rm M} = 0.3$, $\Omega_\Lambda = 0.7$ and a \cite{chabrier2003} initial mass function (IMF), to derive both star-formation rates and stellar masses.

\section{Data \label{SEC:data}}

The data we use in this paper consist of ALMA observations of \hyde obtained in two different programs. The first data set was obtained in the Director's Discretionary Time (DDT) program 2015.A.00026.S (PI: Schreiber) to measure the \cii emission; the galaxy was observed in band 8 (TDM correlator, covering $401.05$--$416.68\,{\rm GHz}$ with four $1.875\,{\rm GHz}$ spectral windows at $31.25\,{\rm MHz}$ resolution) for 1.2 hours (on-source), with a synthesized beam size of $0.52$$\times$$0.42\arcsec$ (natural weighting). These observations were already presented in \citetalias{schreiber2018}, and were reprocessed here for homogeneity. The second data set was obtained in the regular call program 2018.1.00216.S (PI: Schreiber) to measure the \nii emission; the galaxy was observed in band 7 (TDM correlator, covering $295.34$--$311.20\,{\rm GHz}$ with four $1.875\,{\rm GHz}$ spectral windows at $31.25\,{\rm MHz}$ resolution) for 0.8 hours (on-source), with a synthesized beam size of $0.29$$\times$$0.25\arcsec$ (natural weighting).

Both data sets were reduced with the same procedure, using the ALMA pipeline to produce dirty images with natural weighting (to maximize $S/N$) and a spectral averaging of three elements (so that the spectral response function is effectively one channel, see Section A.6.1 in the ALMA Proposer's Guide). The pixel size was left to its default value of $0.05\arcsec$ and $0.085\arcsec$ in band 7 and band 8, respectively (corresponding to $0.36$ and $0.61\,{\rm kpc}$), which generously samples the core of the dirty beam; finer pixel sizes would not benefit the quality of the analysis. Since both data sets were observed with the TDM correlators, the final cubes have a spectral resolution of $35$ and $45\,{\rm km/s}$, respectively, which is sufficient to resolve the broad line profiles expected in massive compact galaxies.

Compact continuum emission was clearly detected in both data sets ($S_{\rm peak}/N$ of $46$ and $41$, respectively), which had to be removed prior to analyzing the line emission. To optimize the $S/N$, this was done in the image domain by fitting the best continuum spectral model obtained in \citetalias{schreiber2018} to the spectral data at each pixel of the image, excluding spectral elements with expected line emission. The resulting spectra, extracted at the peak pixel, are shown on \rfig{FIG:specs}. Line maps were finally created by spectrally-averaging the continuum-subtracted cubes over $800\,{\rm km/s}$ around the expected line frequencies at $z=3.7087$ (see \rfig{FIG:specs}), the redshift obtained from the \cii emission in \citetalias{schreiber2018}. The lines were detected with a $S_{\rm peak}/N$ of $17$ and $5$, respectively, and the maps are shown in \rfig{FIG:maps}. Despite these moderate $S_{\rm peak}/N$, the detection of \nii is still highly significant: since we did not fit for the position of the line emission in neither the spatial not the frequency domains, the null hypothesis of a non-detection can be safely rejected.

\section{Spatial profile modeling \label{SEC:model}}

\subsection{Description of the modeling}

\begin{figure}
\begin{center}
\includegraphics[width=0.49\textwidth]{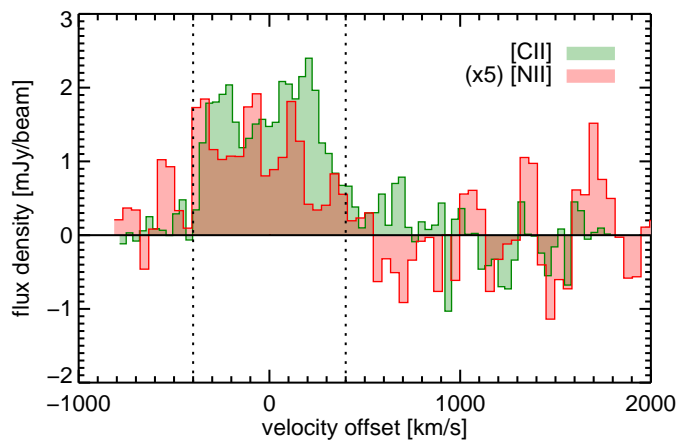}
\end{center}
\caption{\cii (green) and \nii (red) spectra extracted at the peak pixel of the line emission, after continuum subtraction. The \nii spectrum was rescaled upward by a factor five for easier comparison to the \cii spectrum. The integration window used to create the line maps is shown with vertical dotted lines. \label{FIG:specs}}
\end{figure}

\begin{figure}
\begin{center}
\includegraphics[width=0.49\textwidth]{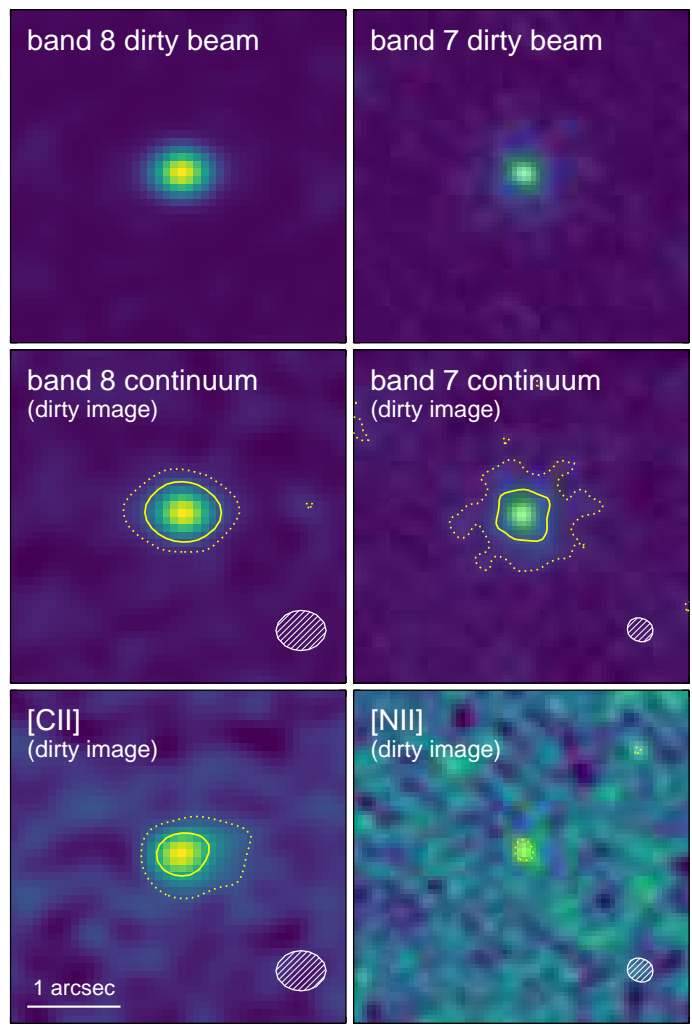}
\end{center}
\caption{Dirty beam (top), continuum (center) and line maps (bottom) of \hyde in band 8 (left) and 7 (right) produced by the ALMA pipeline, without any cleaning applied (``dirty'' images). Line maps are continuum-subtracted and were produced by summing the flux in a $800\,{\rm km/s}$ velocity window centered on the \cii mean velocity. The half-intensity area of the corresponding dirty beams are shown in the bottom-right corner of each panel. Contours shown are $3\sigma$ (dotted line) and $10\sigma$ (solid line). \label{FIG:maps}}
\end{figure}

Using a custom-build software, we then modeled each line map independently in the image domain. We did not attempt to model the maps directly in the visibility domain for two reasons; first, fitting complex profiles other than Gaussians and point sources in the visibility domain is particularly challenging and time-consuming, and second, it has been shown that image-based and visibility-based methods actually provide similar results for ALMA data \citep{hodge2016}. Nevertheless, we double check in \rsec{SEC:method_valid} the result of our modeling against the observed visibilities.

Our model consists of a large grid of two-component exponential profiles. The first ``central'' component was given a small scale length of $0.1$ to $0.5\,{\rm kpc}$ to represent an unresolved central region (\citetalias{schreiber2018} showed the continuum emission extends over a $\sim0.7\,{\rm kpc}$ half-light radius), while the scale length of the second ``extended'' component was allowed to vary between $0.15$ and $3\,{\rm kpc}$, with the constraint that its scale length must be at least $0.05\,{\rm kpc}$ larger than that of the central component\footnote{We tried extending this grid to a maximum of $1.5\,\kpc$ for the ``central'' component and $5\,\kpc$ for the ``extended'' component; this had no significant impact on our flux measurements, suggesting the original grid was large enough.}. We let the flux of each component vary independently and freely, with the sole constraint that the combined light profile must remain positive at all positions. We also varied the axis ratio and the position angle, however to save on computation time we forced the values of both components to match. Finally, the position of both components was fixed to the centroid of the continuum emission. Since the size of the galaxy is small compared to the image pixel size, we generated models with a $\times$$9$ oversampling factor, and a further $\times$$3$ oversampling for the central $3$$\times$$3$ pixels to properly sample the exponential core. This corresponds to an over-sampled pixel size of $0.01$ (band 6) and $0.02\,{\rm kpc}$ (band 8) at the core. We then convolved the model with the ALMA dirty beam, and selected the best-fit model using maximum likelihood estimation, accounting for spatially-correlated noise in the computation of the likelihood (see \rapp{APP:correl}).

To estimate the uncertainty on the model, we used Monte Carlo simulations: the whole procedure was repeated $200$ times on mock images that were created by perturbing the observed images with realistic noise (same amplitude and same covariance as in the real data, see \rapp{APP:noise}), leading to $200$ other ``acceptable'' models from which we estimated confidence intervals using the $16$th and $84$th percentiles (see \rapp{APP:mc}). Because the distributions of the observables derived from this modeling (e.g., the ionized gas fraction) can be strongly non-Gaussian, we carried out all our calculations on each of the $200$ Monte Carlo realizations, and derived confidence intervals for all derived quantities in the same fashion throughout the paper. With $200$ Monte Carlo simulations, $1\sigma$ asymmetric error bars have an accuracy of $10\%$.

\subsection{Validation of the image-domain modeling \label{SEC:method_valid}}

\begin{figure*}
\begin{center}
\includegraphics[width=0.95\textwidth]{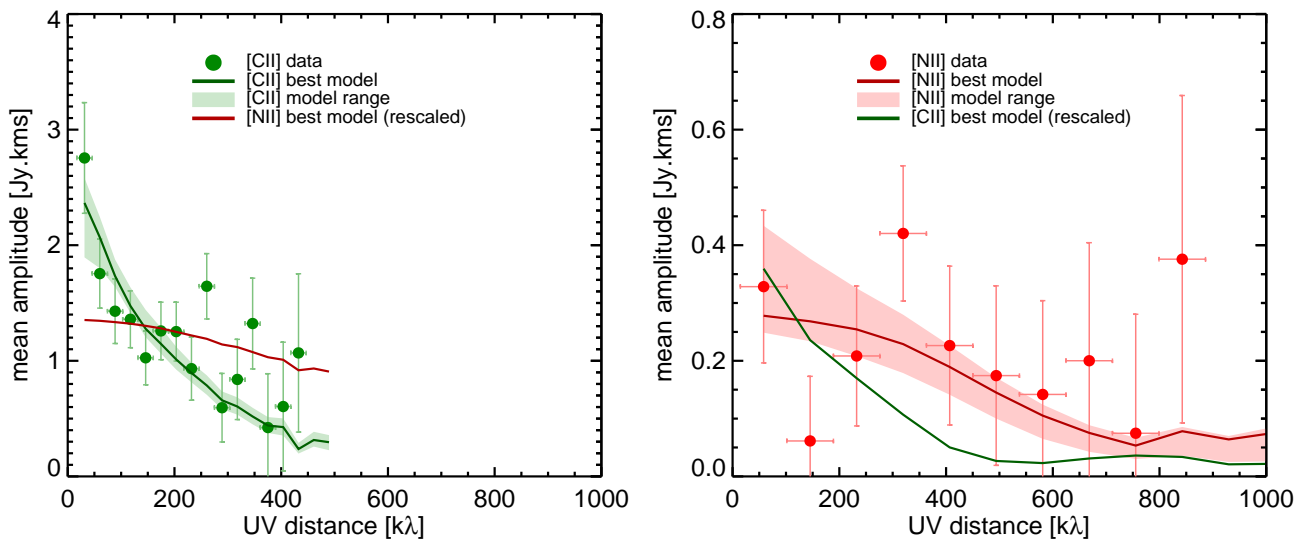}
\end{center}
\caption{Binned amplitude of the ALMA visibilities as a function of the UV distance ($u^2 + v^2$). The \cii visibilities are shown on the left (green), and the \nii visibilities are shown on the right (red). Observed visibilities are displayed as filled circles with error bars; for clarity, the visibilities on long baselines, which have large uncertainties, are not displayed. The best fit models obtained in this paper (from the image-domain analysis) are shown as solid lines (green for \cii, red for \nii). For comparison purposes, on each panel, the model of the other line is also displayed, but re-normalized to fit the observed visibilities. The 1$\sigma$ confidence interval of the model (determined from our Monte Carlo simulations) is shown as a shaded region in the background. \label{FIG:uvamp}}
\end{figure*}

Since we performed all our analysis in the image-domain, we checked that the same trends we detect in this analysis are present in the raw ALMA visibilities. To do so, we used CASA to perform continuum subtraction in the visibility domain using \texttt{uvcontsub} (first order polynomial), excluding channels containing the line in the fit. From the resulting measurement set, we then extracted and averaged the frequency channels covered by the line using \texttt{split}. We note that this data set is not strictly equivalent to the image-domain data we used in our analysis, where the continuum subtraction and frequency averaging were performed in the image domain (with slightly different weights, computed from the image RMS in each channel). The images created from these alternative measurement sets have $10$-$20\%$ larger noise RSM; the $S/N$ will therefore be lower than in our main analysis, but this will be sufficient for our purpose.

We finally extracted the visibilities using \texttt{ms.getdata()} and manually performed the remaining averaging (polarization) and the binning by $(u,v)$ distance. The real and imaginary parts of the visibilities were averaged separately in each bin, using the weights in the measurement set, and combined to form the binned amplitude. The uncertainties on the binned amplitudes were estimated from the weighted standard deviation of the data in the bin, and scaled down by the square root of the number of points.

In parallel, we also produced mock visibilities for our best fit and Monte Carlo models and binned them in a similar way. The corresponding visibilities were obtained from the $\times$9 oversampled model images created by our fitting procedure. These visibilities were computed and injected into the real measurement set using the CASA task \texttt{ft}, which generates mock measurement sets with the exact same $(u,v)$ coverage and weights as the real data. To correct for any discrepancy the data reduction between these visibilities and the images from which we derived the models, we applied a global rescaling factor to our best-fit model amplitudes to best match the observed visibilities.

The binned amplitudes are shown in \rfig{FIG:uvamp}. These figures show the \cii emission is clearly resolved. The \nii emission is more noisy, but appears more compact and consistent with being unresolved. The best-fit models obtained from the image-domain analysis provide good matches to the visibility data, but seem mutually inconsistent between the two lines. In fact, trying to fit the \cii data with the \nii model, we obtained a worse $\chi^2$ ($22.4$, vs $20.9$ for the best fit \cii model), and similarly for the \nii data and the \cii model ($15.0$, vs $8.0$ for the best fit \nii model). This implies that the two emission lines have different spatial distributions, a point we analyze in greater detail in \rsec{SEC:results}.

\subsection{Aperture fluxes}

We first applied our model to the new band 7 continuum image, which has a higher $S/N$ and sharper angular resolution than the band 8 data used in \citetalias{schreiber2018}, and updated the dust continuum half-light radius to $0.51\pm0.07\,{\rm kpc}$ (major axis), which is consistent with the value obtained from the band 8 data. This value was used to separate the emission into a ``center'' and ``outskirts'' components, as we now describe.

Coming back to the other images (line and continuum maps), we summed the flux of every model (both for the best fit and the $200$ Monte Carlo simulations) in two circular annuli: from $0$ to $0.5\,{\rm kpc}$ for the ``central'' annulus, and from $0.5$ to $3.5\,{\rm kpc}$ for the ``outskirts'' annulus. We note that the central annulus actually contains more than half ($\sim75\%$) of the continuum emission since the model axis ratio is lower than one.

The obtained ``central'' and ``outskirts'' fluxes (for both lines and continua) are the main measurements we use in the rest of this analysis. The other parameters of the model, such as the respective size and flux of both exponential components, are effectively marginalized over and considered unimportant. Aperture fluxes are typically simpler to constrain than other shape-related model parameters, because they tend to be less model-dependent. For example, the flux inside our ``central'' aperture depends only slightly on the adopted size for the central component; if the aperture is larger or comparable in size, it then contains all the flux from that component, no matter how that flux is distributed internally. Furthermore, the total flux in that aperture has a natural upper bound from the data. To illustrate this, we can consider the peak pixel of the image, which for \cii entirely contains our aperture. Because of the convolution with the dirty beam, this pixel contains flux from both inside and outside of our aperture. Yet, even if the relative contribution of one versus the other is uncertain, neither can exceed the observed peak flux.

In this light, it may seem overly complex to introduce a rich two-component model to, in the end, simply measure two aperture fluxes. We stress that this model complexity is in fact crucial to assess the reliability of our measurement. Indeed, the constraining power of our data (owing both to $S/N$ and angular resolution) is not sufficient to allow us to determine the exact shape of the intensity profile at all radii. Therefore, by exploring as wide a range of models as possible -- including numerous models which are barely distinguishable on the ALMA images -- we make sure our uncertainties on the aperture fluxes encompass all credible scenarios allowed by the data.

To test our method and the accuracy of our uncertainties, we applied our full measurement procedure to simulated images with injected sources, as described in \rapp{APP:sims}. We found that our method can recover the total, central, and outskirts fluxes in all our simulated images with no detectable systematic bias. The estimated uncertainties from Monte Carlo simulations were found to correctly capture the noise in our measurements, with at most a $10\%$ under-estimation.

\section{Results \label{SEC:results}}

\subsection{Spatial distribution of the lines}

In \rfig{FIG:fion} (left) we show the surface brightness profile of \hyde inferred from our two-component model for the dust continuum and the two emission lines, \cii and \nii. Although the $S/N$ of the \nii data is low, the angular resolution is sufficient to show that most of the emission is centrally-concentrated, with a profile that is similar to that of the dust continuum. The \cii emission, however, appears more extended (as is commonly found in the literature, see, e.g., \citealt{gullberg2018,rybak2019,tadaki2019}). The intrinsic half-light radii we obtained for the continuum, \cii, and \nii are (respectively) $r_{1/2}=0.51\pm0.07$, $1.2\pm0.3$, and $0.3^{+0.3}_{-0.1}\,{\rm kpc}$ (along the major axis). The lower error bar on the \nii size is limited by the minimum source size in our model grid, $0.1\,\kpc$; the data are consistent with the \nii emission being point-like.

\cite{marti-vidal2012} quantify what we should be able to measure in our data given the angular resolution and $S/N$: based on their Eq.~(7), our data should be able to distinguish, with more than $95\%$ reliability ($2\sigma$), between a point source and an extended profile of size $0.30$, $0.96$, and $0.94\,\kpc$, respectively. The sizes of the continuum and \cii are both larger than these values and are indeed ``measured'' ($r/\sigma_r = 7$ and $4$, respectively), and the size we report for \nii indeed has a $2\sigma$ upper limit of $0.9\,\kpc$. Our measurements are thus consistent with these theoretical expectations.

To further confirm this difference in half-light radii with a simpler independent method, we used the simulated noise maps described in the previous section and injected sources of sizes matching the above values, and convolved with the dirty beam. For each source, we computed its observed radial profile on the noisy image, and located the radius at which the emission falls below half of the peak. Comparing this simulation to the values observed on the real images, we found that the \cii emission is significantly ($3.6\sigma$) larger than the continuum, and that \nii is significantly ($2.4\sigma$) smaller than \cii, while the difference between \nii and the continuum is only marginal ($1.2\sigma$).

This confirms that the difference in size between the \cii and \nii emission is not simply due to noise, and implies the presence of a strong gradient in the $\cii/\nii$ ratio, with inner regions having lower $\cii/\nii$ and therefore a higher proportion of ionized gas.

\begin{table*}
\begin{center}
\caption{\label{TAB:fluxes} Measured line fluxes from the image analysis, corresponding ionized gas fractions and infrared luminosities.}
\begin{tabular}{l|cccccccc}
               & $I_\cii$               & $I_\nii$                  & \cii/\nii             & $\fion$          & $\fism$          & $\log_{10}(\lir/\lsun)$ & $\log_{10}(L_{\rm IR,BC}/\lsun)$ & $\log_{10}(\lcii/\lfir)$ \\
Flux origin    & $\Jykms$               & $\Jykms$                  & $^e$                  & \%               & \%               & $^f$                    &                                  & \\ \hline\hline \\[-0.3cm]
Total$^a$      & $1.73^{+0.13}_{-0.10}$ & $0.227^{+0.053}_{-0.046}$ & $9.8^{+2.6}_{-1.8}$   & $28^{+8}_{-6}$   & $60^{+8}_{-10}$  & $12.03^{+0.14}_{-0.14}$ & $11.52^{+0.17}_{-0.24}$ $^g$     & $-2.89^{+0.23}_{-0.26}$ \\[0.1cm]
Central$^b$    & $0.69^{+0.25}_{-0.19}$ & $0.187^{+0.049}_{-0.039}$ & $4.8^{+2.4}_{-1.7}$   & $58^{+32}_{-20}$ & $84^{+12}_{-15}$ & $11.93^{+0.15}_{-0.16}$ & $11.12^{+0.30}_{-0.69}$          & $-3.17^{+0.26}_{-0.28}$ \\[0.1cm]
Outskirts$^c$  & $1.04^{+0.25}_{-0.27}$ & $0.022^{+0.072}_{-0.022}$ & $60^{+\rm inf}_{-46}$ & $5^{+14}_{-5}$   & $13^{+33}_{-13}$ & $11.39^{+0.13}_{-0.19}$ & $11.26^{+0.21}_{-0.21}$          & $-2.46^{+0.33}_{-0.29}$ \\[0.1cm]
Peak pixel$^d$ & $1.24^{+0.07}_{-0.07}$ & $0.190^{+0.038}_{-0.038}$ & $8.6^{+2.4}_{-1.8}$   & $32^{+11}_{-9}$  & $64^{+12}_{-11}$ & $11.95^{+0.14}_{-0.14}$ & $11.37^{+0.17}_{-0.17}$          & $-2.96^{+0.22}_{-0.23}$ \\[0.1cm] \hline
\end{tabular}
\end{center}
$^a$ Summed flux inside a $3.5\,\kpc$ radius.
$^b$ Summed flux inside a $0.5\,\kpc$ radius.
$^c$ Summed flux between $0.5$ and $3.5\,\kpc$ radius.
$^d$ Flux of the peak pixels on the ALMA image (warning: because of the poorer resolution, this method measures flux on a larger scale for \cii than for \nii).
$^e$ Ratio of the \cii to \nii flux, with fluxes expressed in ${\rm W}/{\rm m}^2$.
$^f$ Total infrared luminosity ($8$--$1000\,\um$), computed from the total $\lir$ estimated in \cite{schreiber2018} scaled by the fraction of \nii continuum flux in each region.
$^g$ Computed as the sum of the central and outskirts luminosities.
\end{table*}

To quantify the gradient in the line ratio, we now turn to the measured fluxes. We list the measured values and their uncertainties in \rtab{TAB:fluxes}, and provide in \rapp{APP:mc} the full observed distributions in the Monte Carlo simulations, which we use to determine and propagate uncertainties throughout this paper. We found no significant flux in $\nii$ beyond our fiducial $0.5\,\kpc$ radius, and conversely, we found less than half of the $\cii$ flux is located inside this radius. This leads to a line ratio in the central region of $\cii_{\rm center}/\nii_{\rm center} = 4.8^{+2.4}_{-1.7}$, and a lower limit on the line ratio in the outskirts of $\cii_{\rm outskirts}/\nii_{\rm outskirts} > 15$ ($1\sigma$ limit).

\subsection{Ionized gas distribution}

\begin{figure*}
\begin{center}
\includegraphics[width=0.49\textwidth]{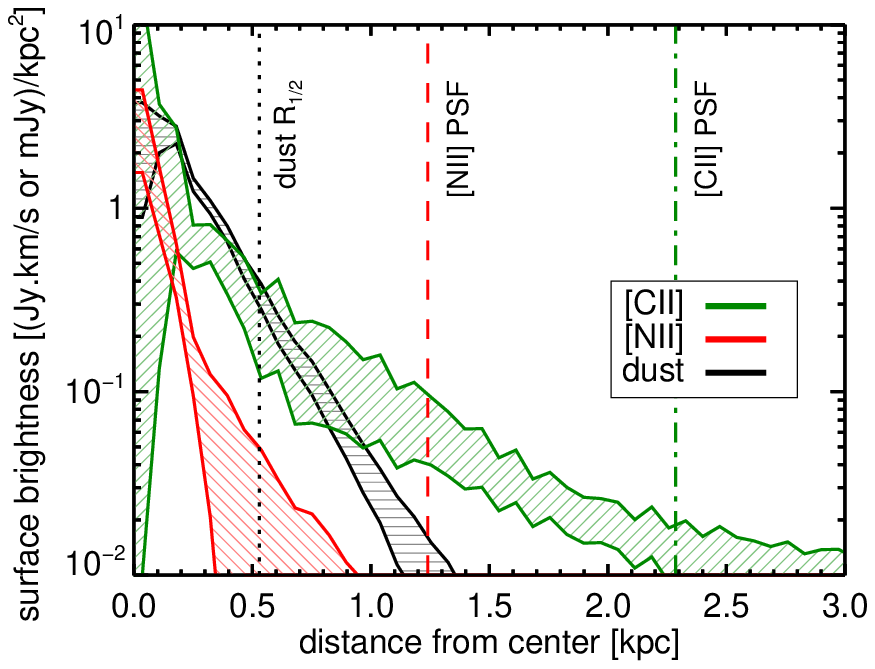}
\includegraphics[width=0.49\textwidth]{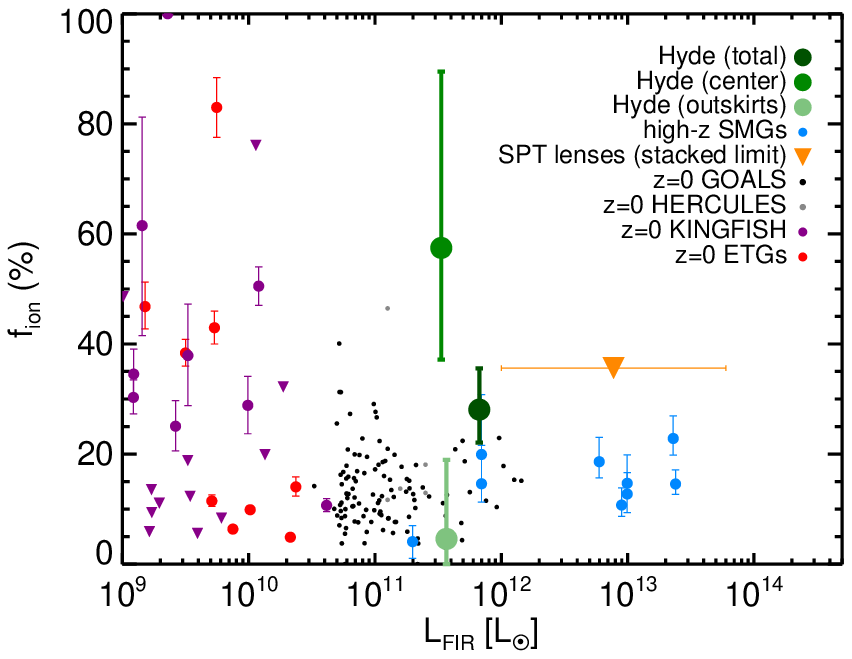}
\end{center}
\caption{{\bf Left:} modeled intrinsic surface brightness profiles of \hyde in the dust continuum (black, horizontal stripes), \cii (green, +45 degrees stripes), and \nii (red, -45 degrees stripes). The dust continuum half-light radius is indicated with a vertical dotted black line. The PSF HWHM of the \cii (resp.~\nii) line map is indicated with vertical dot dashed (resp.~dashed) green (resp.~red) line. The hashed regions show the $1\sigma$ confidence intervals obtained from the $200$ mock noise realizations, and are centered on the best-fit model. {\bf Right:} Ionized gas fraction ($\fion$) determined from the \cii/\nii line ratio in different regions of \hyde (green circles; center, outskirts, and total), compared to other values reported in the literature. For all galaxies, the ionized gas fraction was computed assuming a fixed $\ciion/\niion=2.80\pm0.18$ (see text). We show literature values for local ETGs (red circles) and KINGFISH galaxies (purple circles) from \cite{lapham2017}, local infrared luminous galaxies in GOALS (black circles) from \cite{zhao2016} and \cite{diaz-santos2017} and HERCULES from \cite{rosenberg2015} and \cite{kamenetzky2016}, and high-redshift SMGs from \cite{pavesi2016}. We also show the upper limit of \cite{zhang2018} obtained by stacking high-redshift lensed SMGs (downward-pointing orange triangle).\label{FIG:fion}}
\end{figure*}

Using the fluxes estimated in the previous section, we estimated the fraction of \cii emission associated with ionized gas (or ``ionized gas fraction'' for short), $\fion$, following a method similar to \cite{diaz-santos2017} and \cite{lapham2017}:
\begin{align}
\fion = \frac{\ciion/\niion}{\cii/\nii}\,.
\end{align}
Here we adopt a fixed $\ciion/\niion=2.80\pm0.18$, which is the mean value found by \cite{lapham2017}. In principle this ratio has a weak dependence on the electron density, which can be measured using the $\nii_{122}/\nii_{205}$ ratio, but the $\nii_{122}$ line is not observable for \hyde due to poor atmospheric transmission at that redshifted wavelength. If \hyde turns out to have an unusual electron density, based on the modeling of \cite{lapham2017} its $\ciion/\niion$ could only be higher, up to $\ciion/\niion=4$, which would only increase the value of $\fion$ we infer from the data.

In \rfig{FIG:fion} (right), we display the ionized gas fraction of \hyde in the ``central'' and ``outskirts'' annuli (which we recall correspond to inside and outside of the dust half-light radius, respectively). We find a clear difference between the two regions. The outskirts of the galaxy has a low $\fion$ of $0$--$19\%$, which is typical of infrared-luminous local galaxies and distant SMGs \citep{pavesi2016,diaz-santos2017}. Likewise, the $\lcii/\lfir$ ratio in the outskirts, $\log_{10}(\lcii/\lfir) = -2.46^{+0.33}_{-0.29}$, is similar to that observed in other distant galaxies of this luminosity \citep[e.g.,][]{capak2015}. This suggests that the majority of the \cii emission in the outskirts is associated with cold star-forming gas.

On the other hand, the center of the galaxy has a high\footnote{As a sanity check, we also computed the $\fion$ using the peak fluxes only; these peak fluxes are not model-dependent, but they will contain flux from both inside and outside of our central aperture; we still found a high $\fion$ of $23$--$43$\%, see \rtab{TAB:fluxes}.} $\fion$ of $38$--$90\%$, which has never been observed in a distant SMG and is rarely observed in local infrared-luminous galaxies. Finding such high $\fion$ is instead not uncommon in more normal or quiescent local galaxies, such as main-sequence galaxies and ETGs \citep{lapham2017}. It should be noted, however, that $\fion$ or $\cii/\nii$ values from the literature are typically only quoted for the galaxies as a whole, and few studies prior to this one have attempted to separate the emission from the center and outskirts (see, e.g., \citealt{parkin2013} where a $\fion$ gradient was found in M51). It is possible that higher $\fion$ values could also be found in the center of some SMGs and other IR-luminous galaxies if they were observed with a sufficient angular resolution. This would be an interesting avenue for future observations, in particular as it could allow finding analogs to \hyde in the local Universe, which would provide invaluable insights on the physical processes at play.

The $\log_{10}(\lcii/\lfir) = -3.17^{+0.26}_{-0.28}$ in the center of \hyde is also a factor five lower than in the outskirts, which is expected if the gas is predominantly ionized \citep{diaz-santos2017}. This supports the hypothesis that the majority of the \cii emitting gas in the galaxy center is ionized and not star-forming, and therefore that an energy source other than on-going star formation is significantly contributing to the energy budget in the center of the galaxy. If \hyde is indeed on the path to quenching, this would strongly suggest an inside-out quenching channel.

We cannot determine with certainty the nature of this central ionizing source with the available data, but we can propose two likely candidates. First, the source could be a compact population of intermediate-age stars born in a recent past. Indeed, \hyde is believed to contain about $4\times10^{10}\,\mstar$ of stars within its half light radius (\citetalias{schreiber2018}), and a \cite{bruzual2003} single stellar population of this mass should generate an \nii-ionizing flux with total energy $\gtrsim 10^8\,\lsun$, even at advanced ages of several hundred million years, which is larger than the observed central \nii luminosity, $(6.5^{+1.7}_{-1.4})\times10^7\,\lsun$. However, hydrogen and carbon and will also consume part of this ionizing flux; determining whether this is indeed a viable hypothesis would require dedicated modeling, a measure of the gas metallicity, and a better understanding of the geometry of the system, which are all presently lacking.

Second, the ionizing source could be an AGN. Given that none of the available data indicate the presence of an AGN in this galaxy ($L_X < 10^{44}\,{\rm erg/s}$; \citealt{civano2016}, $L_{1.4\,\rm GHz} < 10^{24}\,{\rm W/Hz}$; \citealt{smoli2017-a}), this may seem less likely, yet a weak or obscured AGN cannot be ruled out. The presence of an AGN would make the interpretation of the observed line ratios more difficult, and could invalidate some of the assumptions and finer calculations presented in the following section, which only account for stellar ionization flux. Yet, the implication on the galaxy's $\sfr$ estimate would follow a similar path: if an AGN exists in the center of this galaxy, with a luminosity large enough to affect the continuum and line emission, it would necessarily contribute to the observed $\lir$. This, in turn, would imply that $\sfr$ estimates using $\lir$ would be biased high. This would include our initial estimate from \citetalias{schreiber2018}, which already tentatively placed the galaxy below the galaxy main sequence. If an AGN is indeed present, and given the large amount of dust and the compactness of the galaxy, the AGN radiation could be trapped \citep[e.g.,][]{costa2018} and couple efficiently with the gas to suppress star formation.

\subsection{Revised star formation rate \label{SEC:revised_sfr}}

\begin{figure}
\begin{center}
\includegraphics[width=0.5\textwidth]{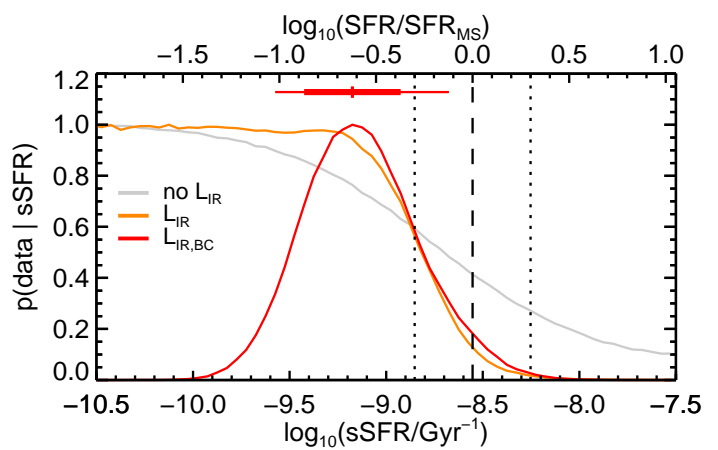}
\end{center}
\caption{Likelihood of the photometric data for \hyde as a function of $\ssfr = \sfr/\mstar$. The likelihood is shown without any constrain on the dust luminosity (gray), using the total infrared luminosity $\lir$ (orange), or using the birthcloud infrared luminosity $L_{\rm IR,BC}$ (red). The horizontal red lines show the $68\%$ (thick) and $90\%$ (thin) confidence intervals. The vertical lines indicate the locus of the galaxy main-sequence and its scatter from \cite{schreiber2017}. \label{FIG:ssfr}}
\end{figure}

\begin{figure}
\begin{center}
\includegraphics[width=0.5\textwidth]{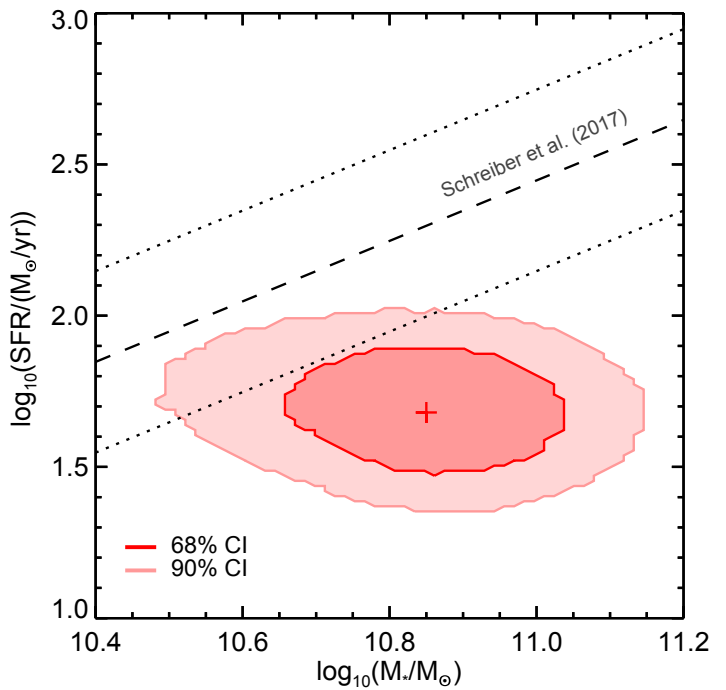}
\end{center}
\caption{Updated location of \hyde with respect to the galaxy main-sequence. The dashed (resp., dotted) line indicate the locus of the galaxy main-sequence (resp., its scatter) from \cite{schreiber2017}. The dark (resp., light) red region shows the final 68\% (resp., 90\%) confidence region for \hyde obtained in this paper. The red cross indicates the location of our best-fit solution. \label{FIG:sfr_mstar}}
\end{figure}

We now use the characterization of the ionizing emission presented in the previous section to provide a refined estimate of the galaxy's $\sfr$. We do so by estimating the infrared luminosity produced in birth clouds by young stars, and feed this luminosity into our SED modeling as a prior.

The total infrared luminosity, $\lir$, is a very useful observable in SED modeling, as it offers an independent constraint on the obscured luminosity of a galaxy. This is particularly crucial to constrain the $\sfr$. However, like \cii, the $\lir$ can originate from different regions in the galaxy. Here we will consider two main contributors: birthclouds (BC), which are heated exclusively by stars younger than $10\,\Myr$ \citep{charlot2000}, and the inter-stellar medium (ISM), which is heated by the older stars \citep[e.g.,][]{dacunha2008}. We note the corresponding infrared luminosities $L_{\rm IR,BC}$ and $L_{\rm IR,ISM}$, respectively. $L_{\rm IR,BC}$ is the quantity we aim to estimate, as by definition it is most directly tracing the recent $\sfr$.

Next, we need to estimate the fraction of the infrared luminosity produced in the ISM, $f_{\rm IR,ISM} = L_{\rm IR,ISM}/\lir$. Intuitively, we can expect this fraction to be closely related to $\fion$, which we computed earlier, since PDRs are co-located with birthclouds, and since ionized gas is part of the ISM. In what follows we will therefore assume $L_{\rm \cii,ion} = L_{\rm \cii,ISM}$ and $L_{\rm \cii,PDR} = L_{\rm \cii,BC}$.

With these assumptions, we can relate $f_{\rm IR,ISM} = L_{\rm IR,ISM}/\lir$ to $\fion$, using the following relation:
\begin{align}
\frac{1}{f_{\rm IR,ISM}} - 1 = \frac{L_{\rm IR,BC}}{L_{\rm IR,ISM}} &= \frac{L_{\rm IR,BC}}{L_{\cii,\rm BC}}\,\frac{L_{\cii,\rm ISM}}{L_{\rm IR,ISM}}\,\frac{L_{\cii,\rm BC}}{L_{\cii,\rm ISM}} \nonumber \\
&=\frac{(\cii/{\rm IR})_{\rm ISM}}{(\cii/{\rm IR})_{\rm BC}}\,\left(\frac{1}{f_{\rm \cii,ion}} - 1\right) \nonumber \\
&= \alpha\,\left(\frac{1}{f_{\rm \cii,ion}} - 1\right) \label{EQ:alpha}
\end{align}

We can determine $\alpha$ empirically using a set of reference galaxies for which we can estimate both $\fion$ and $\fism$. Here we used the local infrared-luminous galaxies from GOALS \citep{diaz-santos2017}, which have direct measurements of \cii and \nii from \herschel and an average of $\fion=12.6\pm0.7\%$, similar to high-z SMGs. Based on the short gas depletion times typically observed for starbursting galaxies (e.g., \citealt{bethermin2015}), we assumed that the GOALS galaxies were mostly formed in a brief burst lasting $50$ to $200\,{\rm Myr}$\footnote{As shown in \cite{dacunha2010-a}, starbursting galaxies can also contain a component of older stars, which would also contribute to the IR emission. If true, this would increase our estimate of $\alpha$ from the GOALS galaxies, which would in turn decrease the final value of $L_{\rm IR, BC}$ and $\sfr$ for \hyde. Our assumption is therefore conservative.}. We then used \cite{bruzual2003} stellar populations and the uniform \cite{calzetti2000} dust screen to compute the expected fraction of their bolometric luminosity produced by young stars for such star formation histories, $f_{\rm bol, BC}=65\pm5\%$. Young stars were defined as stars younger than $10\,\Myr$, as above. Given the strong attenuation in these galaxies \citep{howell2010}, the bolometric luminosity is practically equal to the infrared luminosity, and this latter fraction can thus be identified as $1-\fism$ for GOALS galaxies. Feeding these estimates back to \req{EQ:alpha} gives $\log_{10}(\alpha)=-0.57\pm0.10$ (in other words, the $\cii/{\rm IR}$ ratio is a factor of four times lower in the ambient ISM than in birthclouds).

With the knowledge of $\fism$, which we can compute independently in the center ($\fism=69$--$96\%$) and outskirts ($\fism=0$--$46\%$) of \hyde, we can estimate the summed infrared luminosity produced in birth clouds for both regions: $\log_{10}(L_{\rm IR, BC}/\lsun) = 11.52 \pm 0.20$. We then ran FAST++\footnote{\url{https://github.com/cschreib/fastpp}} v1.3 to model the multi-wavelength photometry with the same setup as in \citetalias{schreiber2018}, but now using $L_{\rm IR, BC}$ as a prior on the obscured luminosity of stars younger than $10\,\Myr$. Briefly, in \citetalias{schreiber2018} we used a flexible model for the star-formation history, formulated as an exponentially rising $\sfr$ followed by an exponential decline, with a variable time of transition between the two phases and variable exponential timescales.

The outcome is illustrated in \rfigs{FIG:ssfr} and \ref{FIG:sfr_mstar}. The stellar mass was essentially unaffected, but the updated model produced an $\sfr = 50^{+24}_{-18}\,\msun/\yr$ (averaged over the last $10\,\Myr$), which is non-zero. This leads to $\ssfr = 0.71^{+0.58}_{-0.34}\,\Gyr^{-1}.$ With a main-sequence locus at $\ssfr=2.8\,\Gyr^{-1}$ \citep{schreiber2017}, this places the galaxy a factor $4.0$ below the main sequence, with a lower limit of $>$$2.3$ at 68\% confidence (and just $>$$1.3$ at 90\% confidence).

The previous estimate of the $\sfr$ from \citetalias{schreiber2018}, using instead the total $\lir$ as a constraint, only allowed us to obtain an upper limit on the $\sfr$ (see \rfig{FIG:ssfr}). The new data, however, exclude the possibility that \hyde has fully quenched ($\sfr \ll \sfr_{\rm MS}$). While the $90\%$ upper limit on the $\sfr$ is actually unchanged compared to our earlier estimates, we are now able to constrain both sides of the $\sfr$ probability distribution, and we can therefore give more credit to the maximum probability solution and the standard $68\%$ confidence interval. In this light, we can claim that the $\sfr$ is now constrained to intermediate values, a factor of $2$--$10$ lower than the main-sequence level. Accounting for the observational uncertainty and the log-normal main-sequence scatter, the probably of observing $\ssfr < 0.71\,\Gyr^{-1}$ when randomly drawing from the main-sequence distribution is $4\%$.

A factor limiting the strength of this conclusion is that the location of the galaxy main sequence is not yet perfectly determined at $z\sim4$. Here we based our comparison on the estimate from \cite{schreiber2017} since it is also based on ALMA-derived $\sfr$s and with stellar masses derived in a similar way; this mitigates the impact of systematic biases on both $\mstar$ and $\sfr$ estimations, and makes the relative $\sfr$ difference between \hyde and the main sequence more robust. In fact, the most robust comparison would be obtained by applying the exact same method to determine the $\sfr$ for \hyde and main-sequence galaxies. Unfortunately, this cannot be achieved until high-quality \cii and \nii data are available for a representative sample of other normal galaxies. Yet, even setting aside all our $\sfr$ modeling, it is still true that \hyde has a high overall $\nii/\cii$ ratio compared to high-z SMGs, and that it is ``special'' in a number of ways (high attenuation, compact size, etc.). Based on this, we argue that most galaxy are not like \hyde, and therefore that on average the $\sfr$ of typical main-sequence galaxies should be correct.

Nevertheless, we can quantify how the above result depends on the estimated locus and scatter of the main-sequence. If the main-sequence scatter is increased to $0.4\,\dex$ (resp.~$0.5\,\dex$), we find that the probably $P$ of observing \hyde's $\ssfr$ when drawing from the main-sequence distribution increases to $9\%$ (resp.~$13\%$). This does not appear to be supported by observations however, as other references in the literature typically report a lower scatter. For example, if the scatter is decreased to $0.25\,\dex$ (resp.~$0.15\,\dex$; \citealt{speagle2014,pearson2018}), $P$ drops to $2.5\%$ (resp.~$0.5\%$). Similarly, if the main-sequence mean $\ssfr$ is decreased by $0.1\,\dex$ \citep{pearson2018}, $P$ increases to $8\%$. However, most references in the literature using FIR, sub-mm, or radio-based $\sfr$ estimates actually report a similar or higher mean $\ssfr$ at $z\sim4$ and $\log_{10}(\mstar)\sim10.8$; with a difference of $+0.1\,\dex$ \citep{speagle2014}, $+0.02\,\dex$ \citep{tomczak2016}, $+0.03\,\dex$ \citep{bourne2017}, $+0.07\,\dex$ \citep{leslie2020}. Considering an increase of the main-sequence $\ssfr$ by $0.1\,\dex$ would decrease $P$ to $2.3\%$.

\section{Our result in context \label{SEC:discussion}}

\subsection{The transition to quiescence?}

The modeling in the previous section allowed us to obtain a revised $\sfr$ estimate for \hyde, which confirms and refines the earlier estimate from \citetalias{schreiber2018}. This $\sfr$, together with the estimated large stellar mass, would place the galaxy significantly outside of the standard galaxy main-sequence and its scatter (e.g., \citealt{schreiber2017}). This supports the hypothesis that the galaxy does not belong to the main-sequence, and is instead in transition to quiescence.

As in \citetalias{schreiber2018}, we are still not able to determine the future of this galaxy with certainty; although its $\sfr$ at the time of observation does appear to be low, we cannot exclude that this only corresponds to a temporary pause in its activity. At best, we can bring forward two arguments which disfavor (but not disprove) this possibility. Firstly, the galaxy is already among the most massive individual known at high redshift, and sits beyond the knee of the stellar mass function. This implies that, if its $\sfr$ does increase in the future, it cannot do so for very long. Secondly, if high-redshift galaxies commonly experience large (but temporary) fluctuations in their $\sfr$, this would be detectable as an increase in the scatter of the galaxy-main sequence. This does not appear to be the case at least at $z\sim4$ (e.g., \citealt{schreiber2017}).

\subsection{Other evidence}

As pointed out in our introduction, \hyde compiles an unusual combination of observables, when compared to other massive galaxies (or SMGs) at a similar epoch. We expand on these aspects in the following paragraphs, and discuss what additional observations could help confirm or contradict our results.

One particularly unusual combination is that of a lower-than-average dust temperature and a compact geometry. \hyde is indeed unusually compact; in recent sub-millimeter surveys, only 4 to $6\%$ of SMGs turn out to have sizes as small as \hyde (e.g., \citealt{ikarashi2017,gullberg2019}). For massive star-forming galaxies in general, compact sizes are typically observed in starburst galaxies located above the galaxy main-sequence (e.g., \citealt{elbaz2011}). These galaxies also tend to have a higher dust temperature (e.g., \citealt{elbaz2011,magnelli2014,bethermin2015}), which can be seen as a natural consequence of their high luminosity and compact geometry. Based on the very compact size of \hyde, we could have therefore expected to see an enhanced dust temperature, but the opposite is observed.

Indeed, when comparing to the average dust temperature of massive $z\sim4$ galaxies from \cite{schreiber2018-a}, \hyde displays a temperature about $10\,\kelvin$ lower than average. Although dust temperatures are notoriously difficult to measure (e.g., \citealt{casey2012}), in this case we can compare \hyde's temperature to a reference which has been measured with the same method (SED fitting and SED templates) and the same wavelength coverage (from \herschel to ALMA band 7). This eliminates most of the systematics, and allows us to robustly quantity the relative difference.

In an attempt to understand these possibly conflicting observations, we can try to draw a comparison to other known galaxies. Here, we have selected the few known massive galaxies at high-redshift with a similar dust temperature of $\sim$$30\,\kelvin$, and with a known spectroscopic redshift; the latter being required to measure the dust temperature accurately. The two most famous examples include GN20 (\citealt{daddi2009,hodge2013-a,tan2014}) and HDF850.1 \citep{walter2012}. These two galaxies have a half-light radius of $6$--$7\,\kpc$, which is an order of magnitude larger than \hyde. This may on its own explain their lower-than-average dust temperature, although recent results suggest this could also be caused by optically-thick dust, at least for GN20 \citep{cortzen2020}. To our knowledge, the only other reported instance of a low dust temperature combined with a compact geometry can be found in the interacting pair SGP 38326 \citep{oteo2016}; unfortunately this system only has poor wavelength coverage, which renders the temperature uncertain ($\tdust = 33$--$55\,\kelvin$). Although comparably massive to \hyde, these galaxies are also an order of magnitude brighter, hence would fit in the traditional picture of a starburst galaxy.

To date, the case of \hyde therefore appears to be unique. Its low dust temperature could correspond to a softer radiation field (or, equivalently, to a low star formation efficiency), which would match the results in this paper. However, as for GN20, it could also be caused by optically-thick dust. Disentangling the two possibilities would require an alternative measurement of the temperature (e.g., using the CO or [\ion{C}{I}] line ratios; \citealt{cortzen2020}).

\section{Conclusions \label{SEC:conclusion}}

We have obtained new ALMA observations toward the galaxy \hyde to observe its $\nii_{205}$ emission. We showed that the line emission is more concentrated in \nii than in \cii, which implies a gradient in the ionized gas fraction. We found the center of the galaxy to be predominantly ionized, which suggests that young stars in their birth-clouds are not the dominant ionization source in the galaxy center, leaving the room for other sources such as intermediate-age-stars, or an AGN. In contrast, the outskirts are dominated by neutral gas which suggests the \cii emission there is mostly associated with star-forming regions. Using these new insights, we obtained an updated estimate of the galaxy's star formation rate, placing it securely below the galaxy main sequence, hence possibly in transition to quiescence.

These results point toward an ongoing inside-out quenching mechanism, and show that this process may start before the galaxy has fully expelled or consumed its gas reservoirs, as demonstrated by the strong obscuration of this galaxy. This confirms the importance of studying the properties of this galaxy in greater details, to better understand the process of quenching at high redshifts.

Further observations of this system would enable a better understanding of the state of the gas, and of the mechanism responsible for quenching, in particular whether the galaxy hosts an AGN or not. New observations with ALMA are already scheduled, including high-resolution imaging of the dust continuum emission to establish the morphology and geometry of the galaxy, as well as [\ion{C}{I}] for measuring the cold gas mass to study the interplay between the ionized and neutral gas reservoirs. Otherwise, the {\it James Webb Space Telescope} (\jwst) could be used to look at the rest-frame optical lines, in particular \halpha, to better constrain the $\sfr$ of the galaxy and obtain an estimate of the gas-phase metallicity. Unfortunately, even \jwst may not have a sharp enough resolution to resolve the optical emission line profiles, which would then require larger telescopes, such as the European Extremely Large Telescope (E-ELT). Finally, deep low-frequency radio observations, for example with the LOw Frequency ARray (LOFAR), would reveal even moderate AGN activity and thus determine the role of AGNs in the high-redshift quenching process.

To go beyond the case study of this single object and determine the rate of occurrence of these features, similar observations of a sample of massive, obscured galaxies would be required. As illustrated here, finding transitioning galaxies is not an easy task, especially if most of them are strongly obscured and would not be included in the $H$-band selected catalogs produced from {\it Hubble} imaging. Samples of {\it Spizter}-IRAC-selected galaxies may be more adequate (see \citealt{caputi2015,wang2016-a,wang2019}), and in the near future \jwst will hopefully open this search to fainter, more numerous galaxies.

\begin{acknowledgements}

The authors want to thank the two anonymous referees for their comments that clearly improved the consistency, thoroughness, and overall quality of this paper.

CS would like to thank Leah Morabito and Ian Heywood for sharing their expertise on interferometric imaging.

Most of the numerical analysis conducted in this work has been performed using {\tt vif}, a free and open source C++ library for fast and robust numerical astrophysics (\hlink{cschreib.github.io/vif/}).

This paper makes use of the following ALMA data: ADS/JAO.ALMA\#2015.A.00026.S and ADS/JAO.ALMA\#2018.1.00216.S. ALMA is a partnership of ESO (representing its member states), NSF (USA) and NINS (Japan), together with NRC (Canada) and NSC and ASIAA (Taiwan) and KASI (Republic of Korea), in cooperation with the Republic of Chile. The Joint ALMA Observatory is operated by ESO, AUI/NRAO and NAOJ.

GGK acknowledges the support of the Australian Research Council through the Discovery Project DP170103470.
\end{acknowledgements}

\bibliographystyle{aa}
\bibliography{../bbib/full}

\appendix

\section{Simulated noise maps\label{APP:noise}}

To estimate parameter uncertainties, the method we used in this paper is to repeat our measurement on mock images, created by duplicating the true image and adding different realizations of noise to it. In this experiment, the observed noisy image and our best-fit model become the ``truth'', and we can determine how far away our fitting procedure is from this ``truth'' on each mock image. The strength of this method is that it requires no assumption on the model, and no analytical calculations; the parameter probability distribution can be extracted straight away by gathering the fits to all mock images. Furthermore, it automatically takes care of parameters that are strongly correlated with one another, as well as correlated noise. The difficulty is that it strongly depends on the quality of the noise that is injected on top of the ``truth'' image. For uncertainties to be accurate, the noise needs to have the exact same amplitude and covariance matrix as in the real data.

\begin{figure*}
\begin{center}
\includegraphics[width=\textwidth]{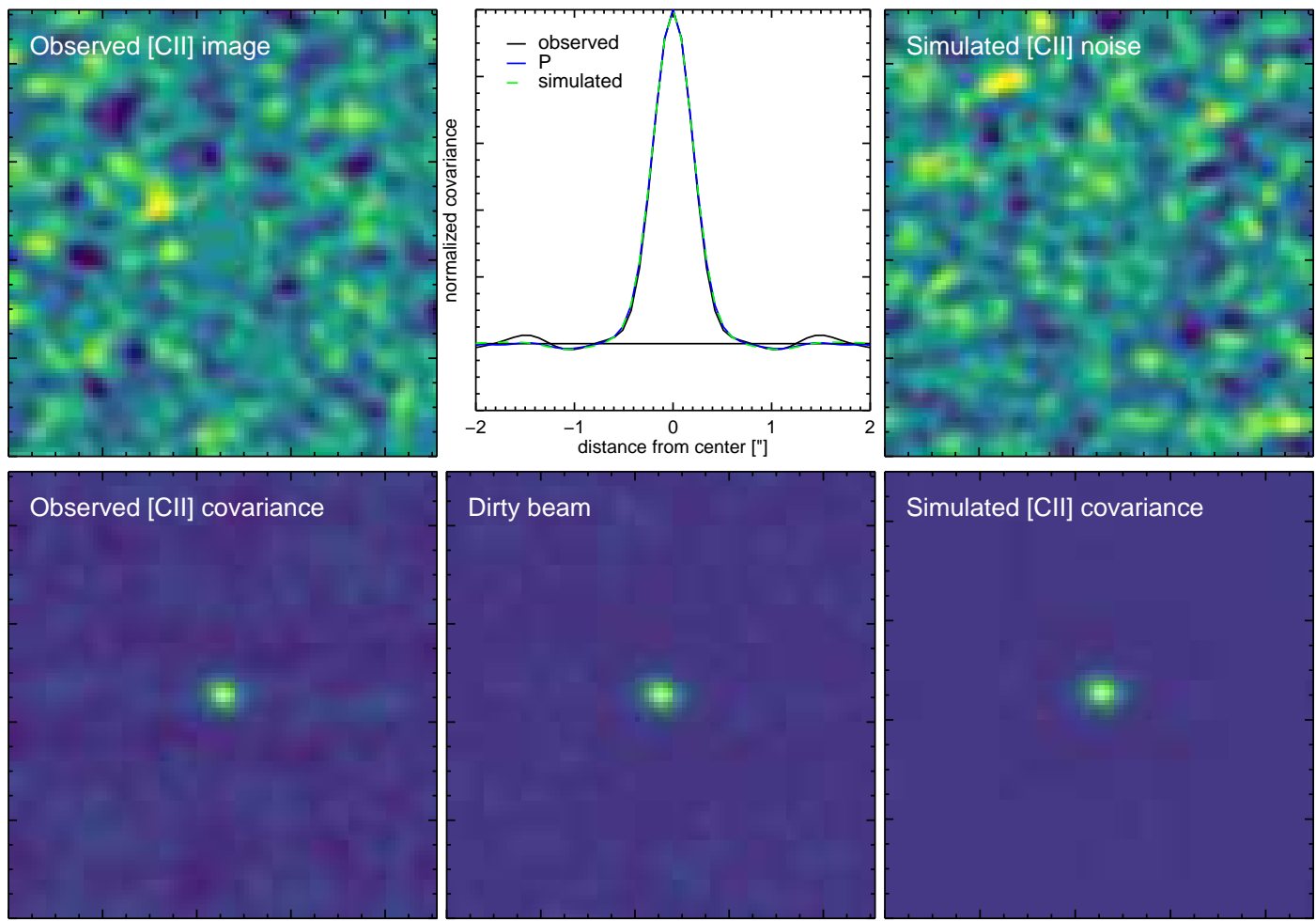}
\end{center}
\caption{Observed and simulated noise maps (top) and their respective auto-correlation functions (bottom). The real \cii map is displayed in the top left corner, with the source masked in the center. For comparison, we also display the \cii dirty beam in the middle of the bottom panel. All the images on each row are displayed with the same color bar. The auto-correlation functions were rescaled prior to display to a peak value of unity. For easier comparison, we also show a circular average of the covariance in shown in the central column of the first row, with the observed (black) and simulated (green dashed) auto-correlation functions, and the dirty-beam profile (blue). \label{FIG:noise_correl}}
\end{figure*}

While in non-interferometric data sets the noise can usually be assumed uncorrelated, this is not the case with our ALMA data; the noise observed in the \cii line map, and its auto-correlation function (ACF), are shown in \rfig{FIG:noise_correl} (left). One can observe that the ACF has a structure very similar to the dirty beam (see \rfig{FIG:noise_correl}, center), and we demonstrate in \rapp{APP:beam_covar} that this is indeed expected when the data is imaged with natural weighting.

This makes it easy to reproduce noise with a similar covariance. Indeed, if a uniform uncorrelated random noise is convolved with a two-dimensional kernel $K$, its ACF will be $C = K\otimes K$, or in the Fourier domain, $\hat{C} = \hat{K}^2$ (where ``hat'' symbolizes the Fourier transform). If we set $C = P$, where $P$ is the dirty beam (see \rapp{APP:beam_covar}), we have:
\begin{align}
 \hat{K} = \sqrt{\hat{P}}. \label{EQ:kernel}
\end{align}
Since the Fourier transform introduces aliasing artifacts on the edges of $K$, it is desirable to use a dirty beam image $P$ that is significantly larger than the final dimensions of the noise map. This can be achieved by padding. To stabilize this further, and since we are mostly interested in preserving the ``core'' of the covariance, we smoothed out the transition to the edges of the image by multiplying the dirty beam image with a broad Gaussian of FWHM $4\arcsec$.

We used this empirical convolution kernel $K$ to generate new correlated noise realizations, shown in \rfig{FIG:noise_correl} (right). The resulting noise maps are visually similar to the real noise map, and the core of the covariance is well reproduced. We note that, since the dirty beam side lobes in our data have a relatively low amplitude, very similar results could have been achieved by simply using a Gaussian kernel of FWHM equal to half that of the dirty beam.

Finally, to set the noise RMS, we simply measured the RMS on the real image and on the simulated noise images, and rescaled the simulated images to match the real observed RMS. Because our target is at the phase center, we did not correct for the primary beam attenuation in the real (or simulated) images, hence the RMS was constant across the entire image.

\section{Auto-correlation function of image-domain noise \label{APP:beam_covar}}

For a generic interferometer, the observed dirty image is the Fourier transform of the weighted and sampled sky visibility:
\begin{align}
I_D(x,y) &\equiv \int A(u,v)\,W(u,v)\,V(u,v)\,\exp\large[-2\pi\,i\,(u\,x + v\,y)\large]\,\dd u\,\dd v\,,
\end{align}
where $A$ is the visibility sampling of the interferometer (a sum of delta functions), $W$ is the imaging weight, and $V$ is the complex visibility. Without loss of generality and for simplicity of notation, we drop the time, frequency, and polarization dependence in all quantities.

The dirty beam is the response to a point source, which has constant visibilities in the Fourier domain:
\begin{align}
P(x,y) &\propto \int A(u,v)\,W(u,v)\,\exp\large[-2\pi\,i\,(u\,x + v\,y)\large]\,\dd u\,\dd v\,.
\end{align}

If we define the inverse Fourier transform of the dirty image
\begin{align}
V_D(u,v) &\equiv \int I_D(x,y)\,\exp\large[2\pi\,i\,(u\,x + v\,y)\large]\,\dd u\,\dd v \\
 &= A(u,v)\,W(u,v)\,V(u,v)\,,
\end{align}
then by definition, the image auto-correlation function (ACF) is
\begin{align}
C(x,y) &\equiv \int V_D(u,v)\,\bar{V}_D(u,v)\,\exp\large[-2\pi\,i\,(u\,x + v\,y)\large]\,\dd u\,\dd v \\
&= \int A(u,v)\,W(u,v)^2\,V(u,v)\,\bar{V}(u,v)\,\exp\large[-2\pi\,i\,(u\,x + v\,y)\large]\,\dd u\,\dd v\,,
\end{align}
where we used $A^2=A$ since it is a sum of delta functions.

To study the ACF of the noise, we now assume that the sky contains no source and the visibilities are therefore only made of noise of amplitude $\sigma(u,v)$, which we assume is uncorrelated. Thus, $\mean{V\,\bar{V}} = \sigma^2$. If we now compute the expectation value of the ACF, we get
\begin{align}
\mean{C(x,y)} &= \int A(u,v)\,W(u,v)^2\,\sigma(u,v)^2\,\exp\large[-2\pi\,i\,(u\,x + v\,y)\large]\,\dd u\,\dd v\,.
\end{align}

In the case of natural weighting, the weights are chosen as $W \propto 1/\sigma^2$. Injecting this into the above equation, we finally get
\begin{align}
\mean{C(x,y)} &\propto \int A(u,v)\,W(u,v)\,\exp\large[-2\pi\,i\,(u\,x + v\,y)\large]\,\dd u\,\dd v \\
&\propto P(x,y)\,.
\end{align}
Therefore, for images produced with natural weighting, the expectation value of the noise ACF is the dirty beam.

In practice the match is not perfect (see \rfig{FIG:noise_correl}), and this can be explained by at least three possible causes: a) we can only measure the ACF on a finite-sized image, so our estimate of the image ACF is noisy; b) the $\sigma$ that enters in the definition of $W$ is only an estimate of the true visibility uncertainty; and c) the noise in the visibilities can itself be correlated. These potential issues seem to have only a moderate impact however; as demonstrated in \rfig{FIG:noise_correl} (middle, top), in our data the core of the ACF is extremely well reproduced by the dirty beam.

\section{Likelihood for correlated Gaussian noise\label{APP:correl}}

The general formulation of the likelihood for Gaussian correlated noise is:
\begin{align}
\mathcal{L} \propto \exp(-\frac{\chi^2}{2})\,, \quad\text{with}\quad \chi^2 = (\mathbf{d} - \mathbf{m})^t\,\Sigma^{-1}\,(\mathbf{d} - \mathbf{m})\,, \label{EQ:chi2}
\end{align}
where $\mathbf{d}$ is a vector of observed data, $\mathbf{m}$ is a vector of model data, and $\Sigma$ is the data covariance matrix (symmetric and positive definite). For our data (see \rapp{APP:beam_covar}), we have:
\begin{align}
\Sigma_{ij} = \sigma^2\,P(x_i - x_j, y_i - y_j)\,,
\end{align}
where $\sigma$ is the image RMS, $(x_i,y_i)$ are the image coordinates of pixel $i$, and $P(\delta x,\delta y)$ is the value of the dirty beam at an offset position $(\delta x, \delta y)$ from the peak.

In practice, inverting $\Sigma$ is numerically unstable. The best way to evaluate it is to perform a Singular Value Decomposition (SVD) such that $\Sigma = U\lambda\,U^t$, where $U$ is a unitary matrix ($U^{-1} = U^t$) and $\lambda$ is a diagonal matrix whose entries are the singular values $\lambda_i$ sorted by decreasing value. Then we have $\Sigma^{-1} = U\lambda^{-1}\,U^t$. If we define $L^{-1} = (1/\!\sqrt{\lambda})\,U^t$, then \req{EQ:chi2} can be rewritten in a simpler form:
\begin{align}
\chi^2 = (\tilde{\mathbf{d}} - \tilde{\mathbf{m}})^t\,(\tilde{\mathbf{d}} - \tilde{\mathbf{m}}) \label{EQ:chi2_simpler}
\end{align}
where $\tilde{\mathbf{d}} = L^{-1}\,\mathbf{d}$ and $\tilde{\mathbf{m}} = L^{-1}\,\mathbf{m}$. These can be seen as the ``de-correlated'' observation and model vectors, respectively.

The SVD is not sufficient to make the computation of $\Sigma^{-1}$ or $L^{-1}$ stable. In all cases, the matrix $\lambda$ contains small entries that cannot be inverted safely. A workaround is to truncate the matrix $(1/\!\sqrt{\lambda})$ by setting to zero the inverse of those singular values that are smaller than some chosen threshold. The resulting matrix inverse is then a ``pseudo-inverse'' (i.e., an approximation of the true inverse), but it is usually better behaved.

The method we adopted to choose the threshold is the following. We defined the normalized singular values $\bar{\lambda}_i = \lambda_i/\lambda_0 \le 1$, and we created a logarithmic grid of $100$ threshold values $v$ ranging from ${\rm min}(\bar{\lambda})$ to $0.1$, such that all $\bar{\lambda}_i < v$ were removed from the inverse. For each value of $v$, we evaluated the corresponding $L^{-1}$, and used it to ``de-correlate'' one of our noise realization from the simulations (this noisy image did not contain any source). We computed the covariance matrix of the resulting image, normalized it to unit diagonal, and defined the metric $k$ as the sum the absolute value of entries with $|x_i - x_j| \le 1$ and $|y_i - y_j| \le 1$. We then picked the value of $v$ which minimized $k$, or in other words, the value which produced an image with the lowest noise covariance. Experiments showed that $k$ is large when $v$ is too small, as the matrix inverse is unstable and the resulting images are degraded. On the other hand, $k$ is also large when $v$ is too large, as this leads to a poorer approximation of the inverse which leaves more correlated noise.

We note that, while \req{EQ:chi2} is formally the correct expression to compute the likelihood for our data, we obtained very similar results when we used the simpler, standard expression for uncorrelated noise, $\chi^2 = \sum_i\,[(d_i - m_i)/\sigma]^2$. This is because we used Monte Carlo simulations to determine the confidence intervals rather than relying on the shape of the likelihood.

\section{Input/output simulations \label{APP:sims}}

\begin{figure*}
\begin{center}
\includegraphics[width=\textwidth]{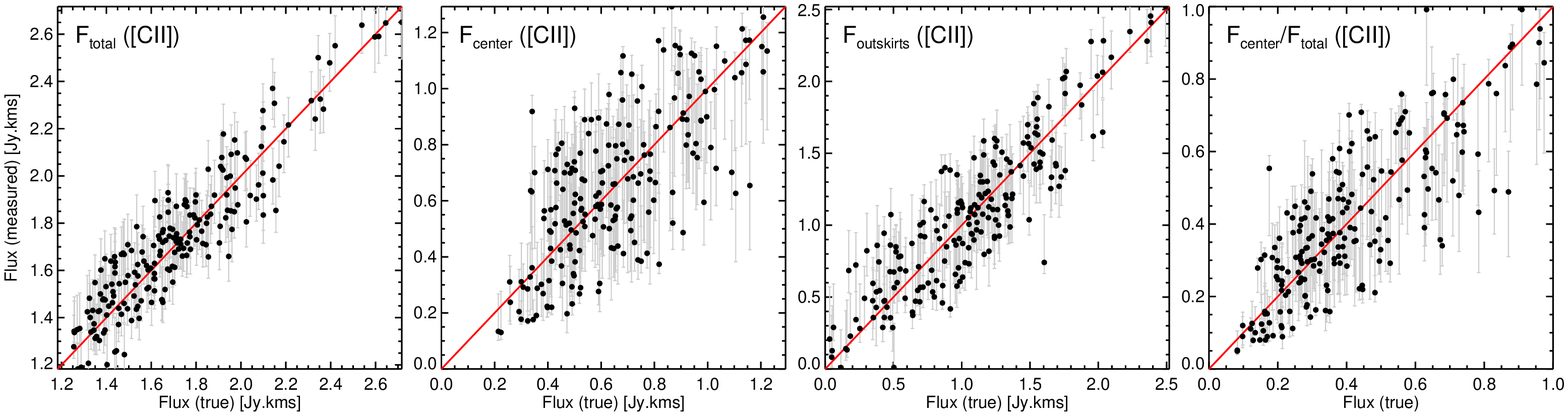}
\includegraphics[width=\textwidth]{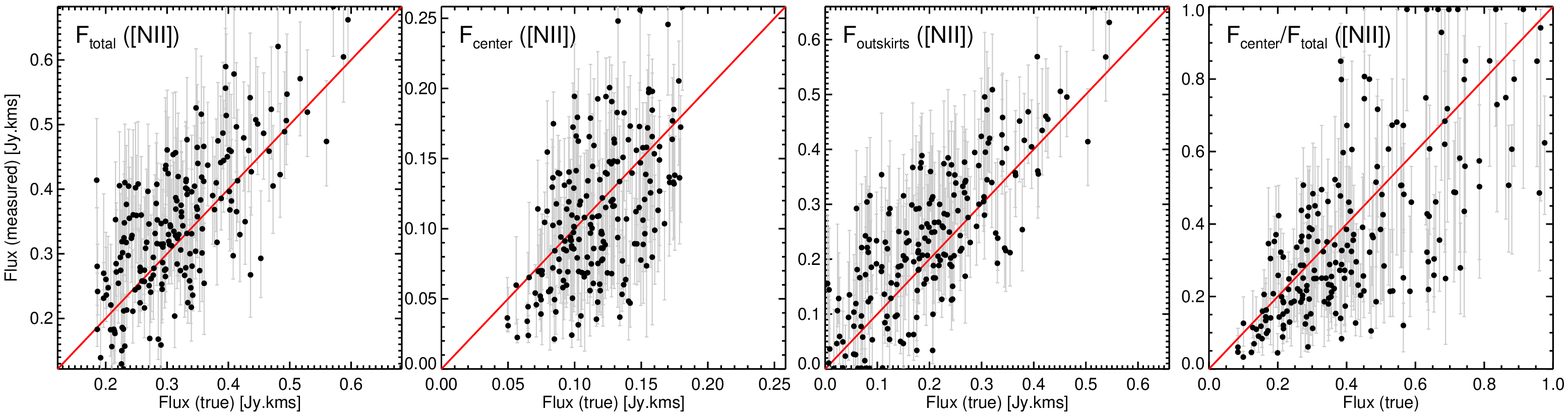}
\end{center}
\caption{Outcome of the input/output analysis of our fitting procedure, modeling sources of known profiles in mock images. The simulations matching the \cii and \nii maps are shown at the top and bottom, respectively. From left to right, we show how our method recovers the total flux, the flux inside the center, the flux in the outskirts, and the ratio of center-to-total (with ``center'' and ``outskirts'' as defined in the main text). The red line is the line of perfect agreement. \label{FIG:sims}}
\end{figure*}

To test the accuracy of our profile-fitting method, we performed an input/output analysis, where we placed sources of known light profiles in simulated noise maps devoid of sources, and tried to recover their profile with our method. The input sources were modeled as two-component exponential disks, with a small ``core'' component and an ``extended'' component, as is assumed in our model. The size of the ``core'' component was chosen randomly in the range $r_{\rm core}=0.1$--$0.5\,{\rm kpc}$, and the size of the extended component was chosen randomly in the range $r_{\rm extended} = r_{\rm core} + 0.2$--$2.2\,{\rm kpc}$. The relative flux of these components was chosen uniformly between 0\% (all flux in extended component) to 100\% (all flux in core component). The axis ratio was chosen randomly between $0.4$ to $1.0$, and the position angle could take any value. Finally, the peak flux of each source was matched to the observed peak flux in our real images. As for our real data, we then computed the true fluxes inside and outside of a $0.5\,{\kpc}$ aperture to obtain the ``central'' and ``outskirts'' fluxes. We note here that, since the \nii image has a sharp angular resolution, the step of fixing the peak flux in the simulations has a strong impact on the true flux in the central aperture, as can be seen in \rfig{FIG:sims} where the range of $F_{\rm center}$ is limited. This is less true for \cii, where the coarser PSF means that the central pixel can contain more flux from the extended component.

The mock sources were injected directly in the image domain, by convolving the source's known light profile with the image dirty beam at $\times$$9$ oversampling. To test the accuracy of this source injection method, we also used CASA to create simulated visibilities for each mock source using \texttt{simobserve} (without thermal noise), and produced corresponding dirty images using \texttt{clean} with the same (oversampled) cell size. The resulting mock images looked identical to those produced by convolution with the dirty beam. To quantify this, for each mock source we computed the second moment ($R^2$) of the images produced by the two methods (visibilities imaging, and image convolution). We found a maximum relative error on $R^2$ of only $3\times10^{-5}$, i.e., close to numerical noise, with no dependence on the size of the mock sources. This implies that injecting directly in the image domain is an accurate source injection method, which is also much faster than creating mock visibilities and imaging them. We caution that this property only holds because we work with dirty images; such images can be described as a Fourier transform of an incomplete $(u,v)$ plane, but cleaned images cannot.

We repeated this procedure with 200 different mock sources, and for each source we replicated the noise level, noise covariance matrix, peak flux, and dirty beam of the \cii, \nii, and continuum maps separately, to test the impact of the varying $S/N$ and angular resolutions encountered in this work. Our fitting procedure was then applied to each of these mock sources, and the recovered profiles were compared to the truth. Since these simulations are accurate mocks of the real images, they allow us to reproduce exactly the measurements we perform in this paper, so we can study any bias arising from our method.

The outcome of this analysis, illustrated in \rfig{FIG:sims} for \cii and \nii, shows that our modeling method recovers the total, central, and outskirts fluxes with no detectable bias. The measurements are noisy, particularly for the central flux, however the observed scatter is correctly captured by the error bars, which were estimated exactly the same way as for the real data. To quantify this, for each observed quantity $F$ shown in \rfig{FIG:sims}, we computed the normalized residual $\delta_F = (F_{\rm obs} - F_{\rm true}) / \sigma_{\rm obs}$, where $\sigma_{\rm obs}$ is our estimated uncertainty. We then computed the fraction of simulated points with $|\delta_F| < 1$. If our uncertainties are perfectly accurate, and if the noise on the quantity $F$ is roughly Gaussian, this fraction should be equal to $68$\%. Since we only use $200$ points to compute these fractions, the expected statistical error on this number is about $3$\%. We find fractions for \cii (resp.~\nii) of $67$\% (resp.~ $66$\%) for the total flux, $62$\% (resp.~$61$\%) for the flux in the inner $0.5\,\kpc$, and $61$\% (resp.~$62$\%) for the flux in the outskirts. These values are exactly in the expected range for the total flux, but the fractions for the central and outskirts fluxes are slightly lower than $68$\%. For Gaussian noise, this would imply that the uncertainties are under-estimated by at most $10\%$; taking this small correction into account would not affect our conclusions.

\section{Distribution in Monte Carlo simulation\label{APP:mc}}

\begin{figure*}
\begin{center}
\includegraphics[width=\textwidth]{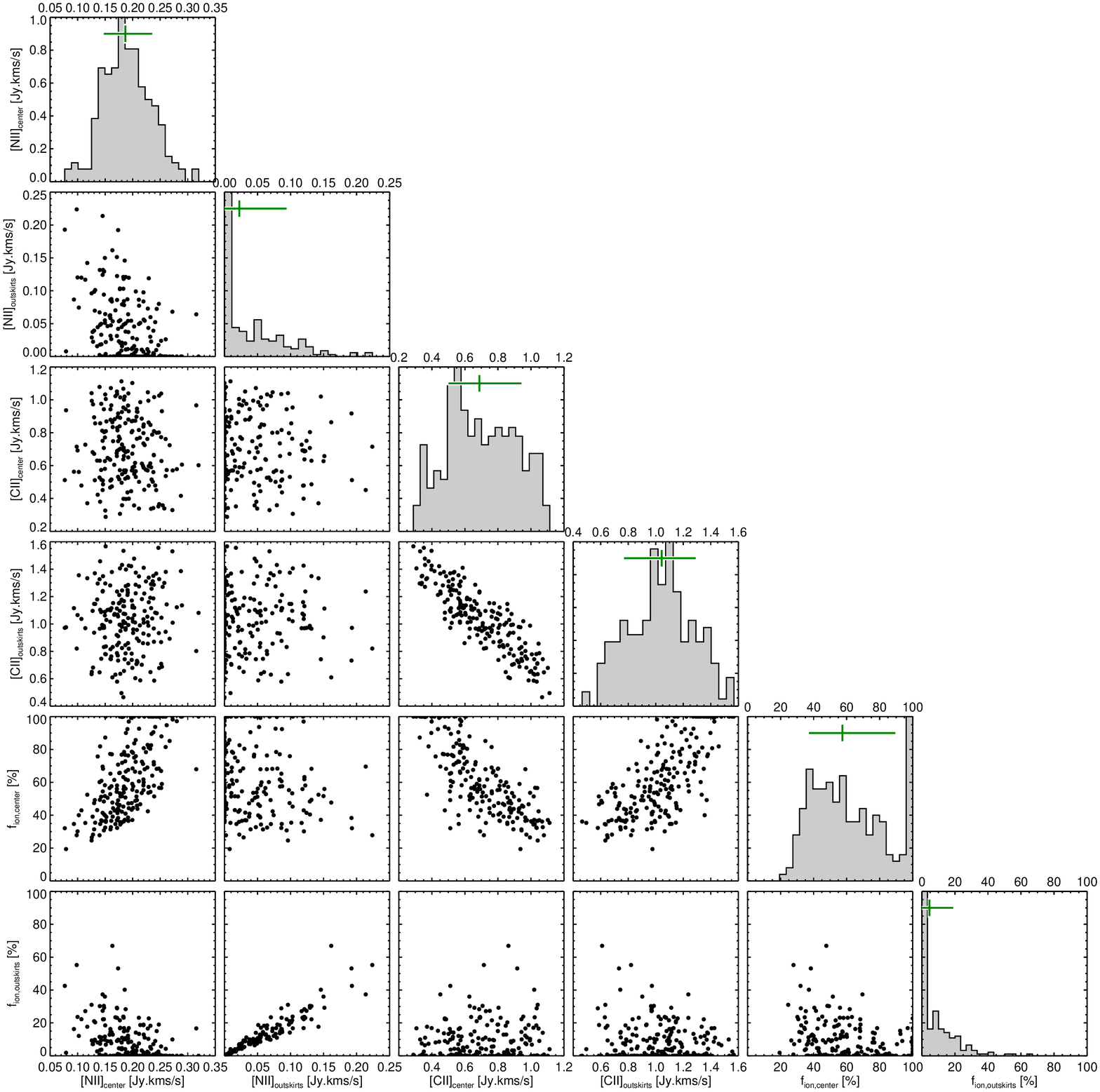}
\end{center}
\caption{Joint distribution of the fluxes, flux ratios, and ionized gas fraction in the Monte Carlo simulations executed for the real images. As is customary, on the diagonal we show the histogram of each quantity; the $y$-axis for these plots is scaled automatically to include the maximum counts in a bin. The green horizontal bars correspond the $16$th and $84$th percentiles, which we report as uncertainties, and the horizontal green bar is the bests-fit value.}
\end{figure*}

\end{document}